# Programmable Motion of Optically Gated Electrically Powered Engineered Microswimmer Robots

Matan Zehavi[1], Ido Rachbuch[2], Sinwook Park[2], Touvia Miloh[2], Orlin D. Velev[3] and Gilad Yossifon[2,4*]

[1]*Faculty of Mechanical Engineering, Technion–Israel Institute of Technology, Technion City 32000, Israel*

[2]*School of Mechanical Engineering, Tel Aviv University Ramat Aviv 69978, Israel*

[3]*Department of Chemical and Biomolecular Engineering, NC State University Raleigh, NC 27695, USA*

[4]*Department of Biomedical Engineering, Tel Aviv University Ramat Aviv 69978, Israel*

* Corresponding author: gyossifon@tauex.ta.ac.il

**Abstract**

Here, we report on a new class active particles capable of dynamically programmable motion powered by electricity. We have implemented physical principles that separate the propulsion and steering mechanisms of active motion using optically activated, patterned, photoresponsive semiconductor coatings on intricate microstructures. Our engineered microswimmer robots employ an induced-charge electro-phoresis (ICEP) mechanism to achieve linear motion and optically modulated electrokinetic propulsion (OMEP) for steering. Optical modulation is achieved by manipulating the polarizability of patterned ZnO semiconductor coating through exposure to light with wavelengths above its bandgap, exploiting the semiconductor's photoconductive properties. Unlike previous methods that rely on changing the direction of optical illumination or spatially controlling narrow optical beams, our approach achieves optical steering under uniform ambient illumination conditions, thereby greatly reducing the complexity of the optical system. The decoupling of propulsion and steering allows for the programming of micromotor trajectories in both open and closed-loop control modes. We anticipate that our findings will pave the way for efficient optically gated control of the trajectory of photoresponsive active particles. Furthermore, they will enable the selective manipulation of specific subgroups of engineered active microparticles with various semiconducting coatings having different band gaps.

**One Sentence Summary:**

This work presents a method for achieving programmable motion in an electrically powered microrobot through optical gating

**Introduction**

The emerging field of self-propelling or active particles represents an exciting and rapidly evolving frontier in both fundamental and applicative science. Active particles are positioned at the intersection of colloid science and the broader realm of active soft matter, where systems are driven into non-equilibrium states through chemical "fuels" or external field energy harvesting. These particles have garnered scientific interest due to their unique ability to convert locally the ambient environmental energy into autonomous translational and/or rotational motion, known as "self-propulsion". Their motion is hard coded into their design, enabling them to draw and dissipate energy via asymmetric patterns and generate local force gradients for propulsion. While externally imposed field gradients can induce particle motion through various phoretic effects such as electrophoresis, dielectrophoresis, magnetophoresis, acoustophoresis and thermophoresis, these mechanisms typically result in en-masse migration along fixed spatial gradients. This contrasts with the individual, independent trajectory freedom of self-propelling particles.

Most active particle systems developed to date rely on a single energy source (e.g., acoustic, electric, magnetic, optical), commonly leading to linear translation in arbitrary directions. Recent studies have also explored the development of active spinners rotating about a central axis (1). However, achieving trajectory control requires the design and development of engineered particles with specialized sizes, shapes, and responsiveness to external stimuli, enabling effective control of motion direction. In the quest for directed motion, particularly for targeted delivery applications, particle propulsion and steering are commonly treated as separate challenges (2). Steering often relies on secondary mechanisms based on physical boundaries (3), hydrodynamic constraints (4), or magnetic guidance (5), providing dynamic trajectory control crucial for targeted delivery but necessitating an additional external magnetic source. Our investigation explores optically steered motion as an alternative to conventional magnetic steering.

Electric fields offer a highly versatile and precise means of propelling active particles, allowing for real-time adjustment of propulsion forces and inter-particle interactions (6). Moreover, simple alterations in electric field frequency can also trigger various electrokinetic effects, such as electrohydrodynamic flows (EHD) (7), induced- charge electrophoresis (ICEP) (8,9), Asymmetric AC Field Electrophoresis (AFEP) (10), self-dielectrophoresis (sDEP) (11,12), and self-electrophoresis by diode rectification (7,8,12–15). While spherical Janus particles have been extensively studied, the complexity of engineered particles introduces a higher level of sophistication. Engineered particles can respond to AC electric fields across multiple frequencies, allowing for multimodal changes in motion direction on demand(16). Recently, the authors explored a new type of supercolloidal particles, termed "microspinners," which are capable of controlled spinning in AC electric fields. These microspinners, characterized by their intricate shapes and discrete metallic patches, convert electrical energy into dynamic motion by leveraging multiple electrokinetic mechanisms across different electric field frequency ranges (1). Understanding the interplay between propulsive effects such as EHD, ICEP, sDEP, and self-electrophoresis is crucial for designing advanced active particles that can harness and channel energy through diverse pathways. In addition to these comprehensive electric-based control capabilities, we aim to further enhance responsiveness by incorporating semiconductor coatings into the particles. Recent studies have shown that microdiodes, powered and controlled by external AC fields, offer new avenues for manipulation (15). Furthermore, the electric polarization of photosensitive semiconductor coatings can be optically regulated via the photoconductive effect (17,18).

We have recently demonstrated the ability to optically tune the mobility of electrically powered Janus particles (JPs) that are half-coated with a photosensitive zinc oxide (ZnO) semiconducting layer with a $SiO_2$ passivation layer. The photoresponse of the ZnO semiconductor increases its electrical conductivity when exposed to light with wavelengths of sufficient photon energy relative to the semiconductor band gap. This effect, which we named optically modulated electrokinetic propulsion (OMEP) (19), can be harnessed to increase the contrast in polarizability between the dielectric and semiconducting hemispheres, resulting in tunable electrokinetic mobility. However, due to the simple axisymmetric structure of the ZnO JP it could only yield linear motion, modulating its velocity magnitude but with no control of direction of motion.

Our objective in this work is to explore how optically activating semiconductor coatings, non-symmetrically patterned on engineered particle surfaces, can decouple steering and linear propulsion mechanisms. Both mechanisms stem from the same underlying induced-charge electro-convective process, but the steering mechanisms is optically gated. By dynamically adjusting the electric polarization of the semiconductor through variations in optical illumination intensity, wavelength, and material selection, we achieved directed motion of complex active structures. Optical steering, resulting from the self-propelling mechanism, differs fundamentally from magnetic steering, which aligns and constrains dipoles parallel to the magnetic field. In contrast, optical steering preserves the autonomy of active particles.

## Results

### *Design, fabrication and overall response of the optically gated engineered active particle*

The microswimmer robot "double H" ('HH') design (Figure 1a) generates propulsion through induced-charge electrokinetic propulsion (ICEP) and steering via optically modulated electrokinetic propulsion (OMEP) mechanisms. The ICEP propulsion is generated over the central metallic (Au) patch that induces a constant electrohydrodynamic effect. For steering, the microswimmer utilizes a Zinc-Oxide (ZnO) semiconductor coating capable of optically altering its polarizability from dielectric to conductive. Unlike our previously studied photoactivated spherical Janus particles (19) which only exhibit linear motion, the geometric symmetry breaking of the 'HH' design allows torque generation upon optical activation enabling controlled steering.

The operational principle of optically gated steering is outlined in Figure 1b. Without UV illumination, the particle undergoes linear translation driven by the ICEP effects from the central section coated with gold (Au), due to the patch symmetric position. When UV illumination is applied, the photosensitive ZnO layer becomes polarizable, initiating electrohydrodynamic motion trough the OMEP effect. This effect induces an additional localized electro-hydrodynamic ejection flow on the ZnO coated section, generating torque on the structure and causing the microdevice to rotate around its vertical axis. These effects are most pronounced at electrical frequencies at or below the RC time constant of the induced electric double layer on both the metallic and photoresponsive patches(12,19,20).

Figure 1c illustrates a particle trajectory wherein rotation is induced by optical gating (blue trace corresponds to UV gated motion), demonstrating the capability for programmable trajectory planning and control. Figure 1d shows the linear and angular velocities for the trajectory depicted in Figure 1c. Switching the UV light ON/OFF reveals that the overall linear velocity slightly increases with UV ON due to the augmented electrohydrodynamic convection in the same direction as on the central Au patch. At this point, the angular velocity undergoes a significant transition from nearly zero to a definitive value. The 'HH' design was preceded by a simpler 'H' design, which related on similar operating principles but had simpler and limited navigation capabilities. See supplementary Figure S4 and Figure S5 for its performance and simulation results.

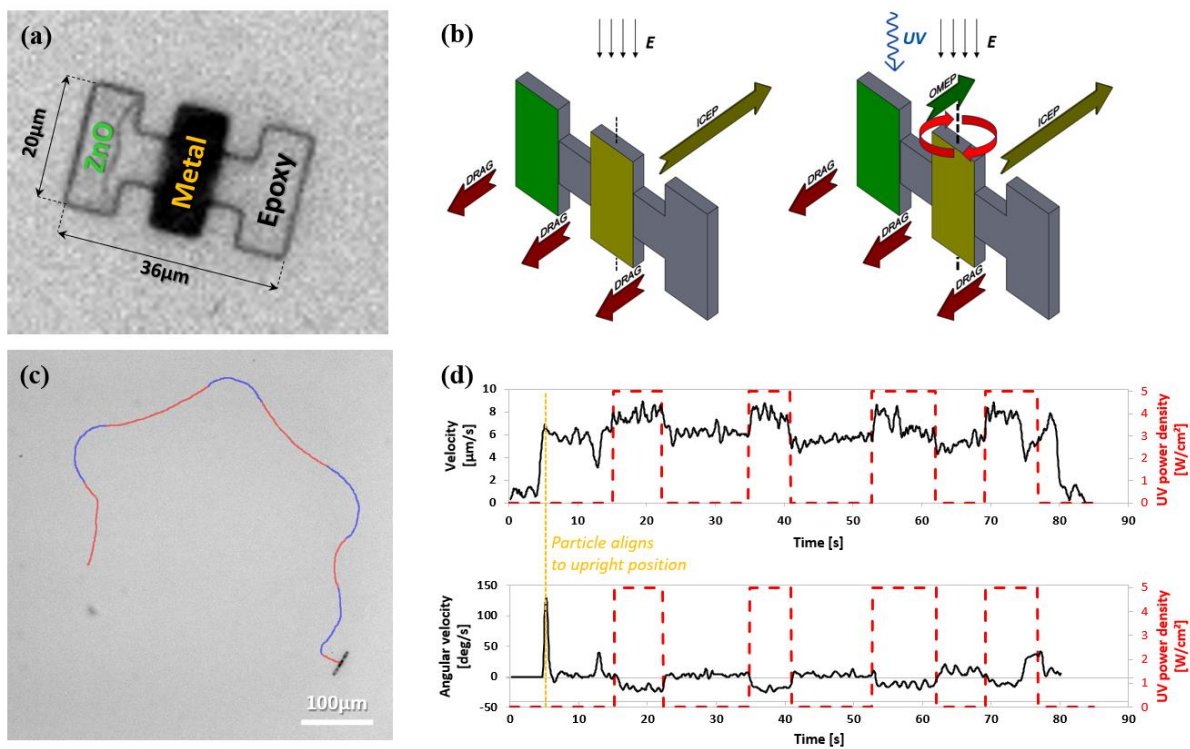

**Figure 1: The 'HH' microswimmer robot and optically gated steering.** (a) A microscope image of the 'HH' type microswimmer. (b) Schematics of the operational principle of steering under an externally applied electric field. On the left, linear motion occurs due to symmetric actuation of the induced-charge electrokinetic (ICEO) effect on the gold-coated surface. On the right, the additional optically modulated electrokinetic propulsion (OMEP) effect, acting on the 'HH' particle's section coated with ZnO generates torque (red arrows) on the microswimmer. (c) Example of the trajectory, and corresponding (d) linear and angular velocities recorded for a 'HH' microswimmer in an open-loop control scenario with UV switched on (blue trace) and off (red trace), under an AC electric field of 2 kHz and 1666 V/cm (see supplementary video SV1).

*Characterization of the electrokinetic propulsion and steering mechanisms*

We conducted comprehensive characterization of the microswimmers to evaluate their performance. To verify that the mechanism propelling the structure is indeed induced-charge electrokinetic propulsion (ICEP) operating on the central metal portion of the swimmer, we conducted a frequency dispersion experiment (Figure 2a) at 1667 Vpp/cm under imaging light only (no UV). The microswimmer demonstrates distinct low-frequency ICEP behavior and a reverse-direction region in the 100 kHz range, which are characteristic of metallo-dielectric Janus and complex active particles (1,12,21). By fixing the frequency at 2 kHz and varying the voltage, we observed that the velocity shows a response proportional to the square of the applied electric field, i.e., $V \propto E^2$ with R² value of 0.98, further reaffirming the ICEP electrokinetic propulsion mechanism (Figure 2b). The OMEP steering mechanism is based on optically tunable conductivity of the ZnO pattern. We measured the ZnO photoconductivity over a test structure (details in the Materials and Methods section) under constant 1V voltage and varying UV illumination power density (Figure 2c). Further, we investigated the impact of UV light application on the linear and angular velocities of the microswimmers at 2 kHz and 1667 Vpp/cm electrical activation, under varying

UV intensities (Figure 2d). The results verify the design's capability to achieve tunable steering without major impact on the linear velocity of the swimmers.

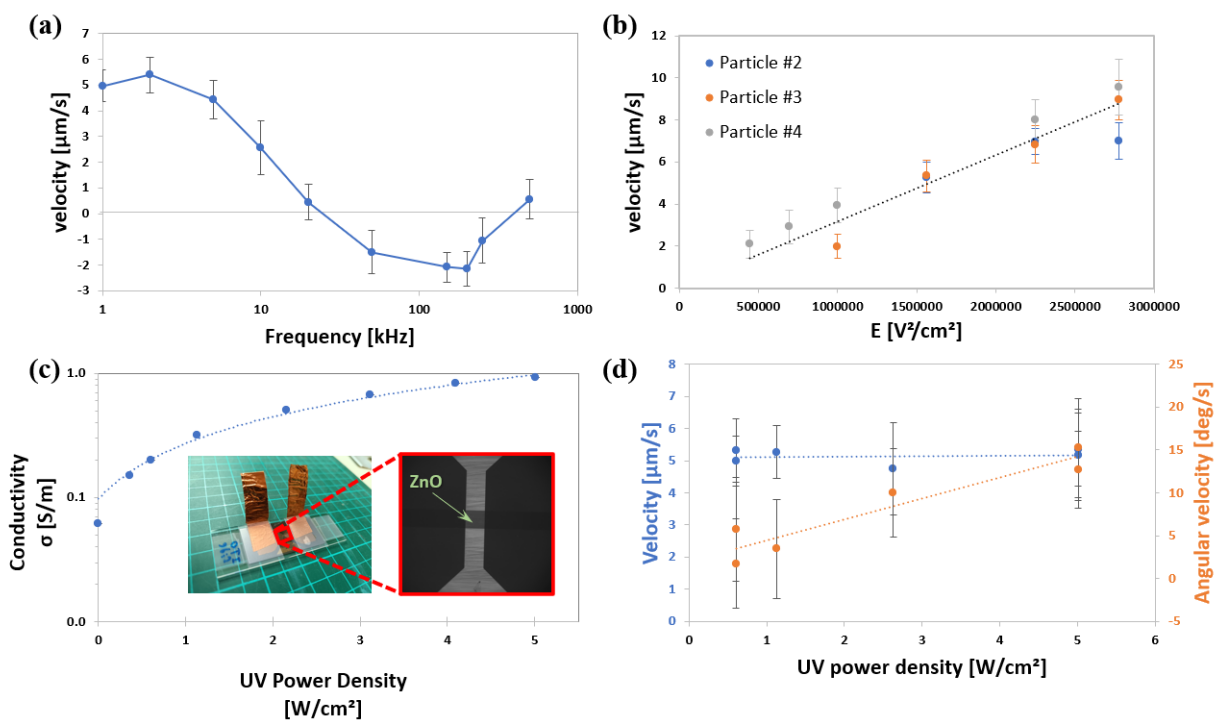

**Figure 2: Characterization of microswimmer material and electrokinetic properties.** (a) Frequency response of a 'HH' microswimmer motility under dark (no UV) conditions for a 1667 V/cm AC electric field with varying frequencies. (b) Motility vs. voltage response at 2 kHz. (c) ZnO layer photoconductivity response to different UV power densities, measured from the photoconductivity test structure (inset) under a constant applied voltage of 1 V. (d) Relationship between linear and angular velocities and varying UV power density at 2 kHz and a 1667 V/cm AC field. Error bars in all plots represent measurement standard deviation.

*Numerical simulations*

Numerical simulations that qualitatively describe the steering mechanism upon switching the UV ON, were performed using the 3D geometry and boundary conditions depicted in Figure *3*a (more details in the Materials and Methods section). Figure *3*b depicts the effect of switching the UV OFF and ON states on the electric potential induced within the microswimmer robot. The ZnO wing exhibits a change in its electronic conductivity when optically activated, thereby altering its induced electrical potential. This change transitions from a linearly varying potential in its dielectric state (UV OFF) to an isopotential in its conductive state (UV ON). The ensuing change in electrohydrodynamic convection is depicted in Figure *3*c indicating the emergence of the OMEP electroconvection over the ZnO wing in addition to ICEP on the metal coated central part of the particle. The velocity streamlines, field vectors and magnitudes are also depicted in Fig. 3d for both the UV ON and OFF states at different plane views. As shown, the induced electro-osmotic flow on the Au and optically gated ZnO patches results in ejection flows from their central region. This ejection flow generates the hydrodynamic force that propels the patch in the opposite direction.

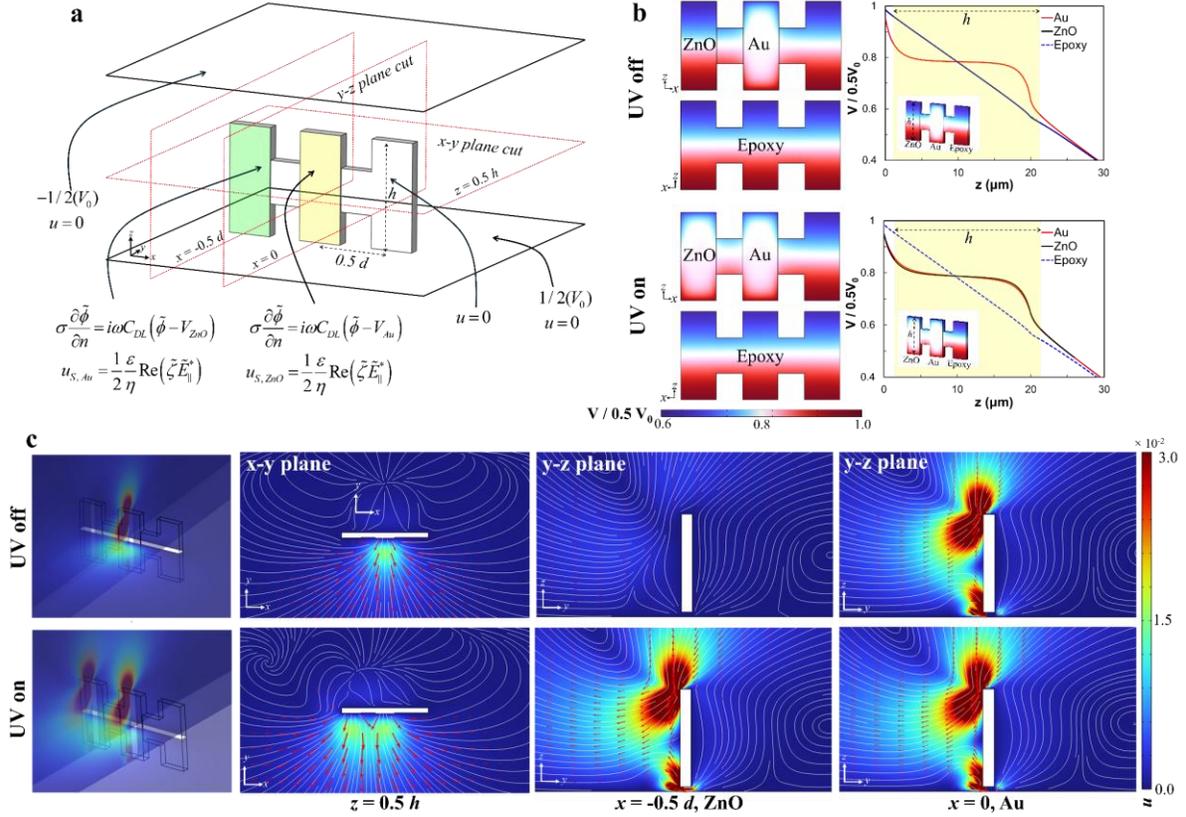

**Figure 3: Numerical simulation results.** (a) Schematics of the 'HH' particle geometry and simulation domain, with boundary conditions. (b) Results of field intensity simulations of the front and back sides of the 'HH' swimmer, showcasing normalized electric potential intensities in UV ON and OFF states. (c) Fluid velocity magnitude plots at the particle surface during both operational states, showing velocity streamlines, vectors (red arrows), and magnitude (color) from various plane views.

*Trajectory planning schemes employing open and closed-loop control*

Here, we demonstrated robust trajectory control of the 'HH' particle by employing both open and closed-loop control-based navigation methods. The open-loop control depicted in Figure 4a shows steering (i.e. turning of the microswimmer) during the UV ON states. The angular velocity of the particle under optical illumination correlates with the UV power density, as depicted in Fig.2d. Directing the particle through different target waypoints and achieving multiple repetitions was challenging due to several factors. Firstly, the self-propelling nature of the micromotor moving autonomously under uniform applied external electric field, makes it sensitive to interactions with debris or other nearby particles. this can result in unexpected abrupt local orientation changes. Secondly, the current OMEP micromotor design cannot turn both ways due to its symmetric geometry and the unidirectional drag force acting on it when the UV is turned off.

Under UV ON conditions, the microswimmer can only turn in the direction dictated by the electrohydrodynamic force acting on its ZnO arm. The closed-loop-controlled paths shown in Figure 4b and 4c exhibit more robust and consistent control for waypoint navigation, compensating for local faults and disturbances to the particle trajectory. This closed-loop control system integrates both high- and low-level control modules (more details in Materials and Methods section). The high-level control primarily involves real-time image analysis for path planning and object tracking, determining the micromotor's orientation, velocity, and trajectory. The low-level control encompasses hardware component interfaces such as the microscope, digital camera, function generator, UV LED source, and the program controller.

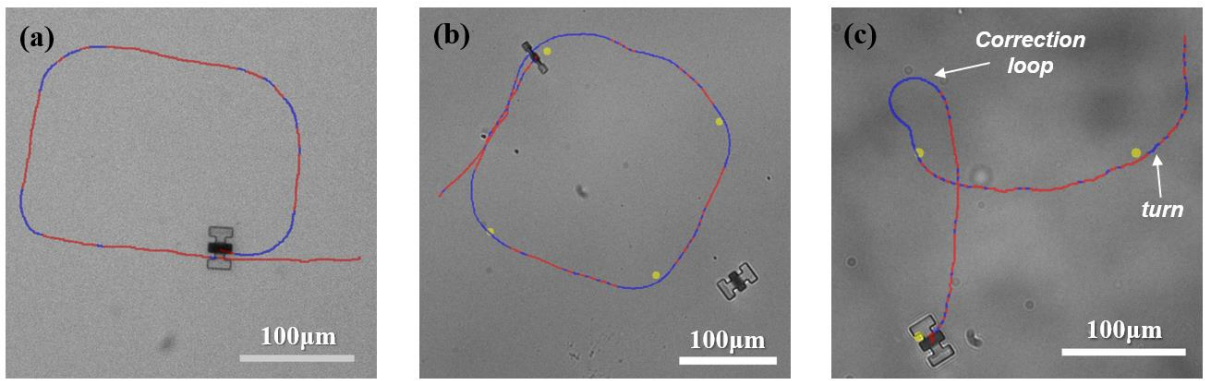

**Figure 4: Trajectory control of the 'HH' microswimmer.** Examples of trajectory control for the 'HH' particles through predetermined waypoints (yellow dots). Red track-lines represent "UV OFF" and blue track-lines represent "UV ON" motion. In all examples, the particle turns clockwise when activated by UV light. (a) Manual (open loop) control of the particle trajectory, drawing a rectangular path (see supplementary video SV2). (b) Automatic (closed loop) microswimmer trajectory control, demonstrating the ability to automatically navigate to predetermined waypoints, showing a tilted square path (supplementary video SV3). (c) Automatic waypoint navigation demonstrating the capability to steer both clockwise and counterclockwise through correction loops (supplementary video SV4), and to compensate for local disturbances.

**Discussion**

We have explored the design and capabilities of a class of optically gated microswimmer robots. Their ability to propel and turn relies on asymmetric design features, using OMEP for steering with a patterned photosensitive semiconductor layer. Both propulsion and steering are based on electrokinetic effects at the metallic (Au) and semiconductor (ZnO) patterned coatings. The main advantage of our OMEP-driven steering approach is the use of a simple modulation of wide-field uniform optical illumination. This stands in stark contrast to other, significantly more complex optical steering mechanisms that require either optical beams with controlled direction of illumination focused beams that precisely illuminate a small section of the particle, or optically patterned trajectory path using digital micromirror device (DMD)(22). While our focus here has been on the 'HH' particle geometry, this selective coating and optical gating method hold promise for various applications including diverse particle geometries, varied photoresponsive coating dimensions, shapes, and semiconductor materials with different band gaps. Numerical simulations complemented the experimental findings and provided a qualitative overview of the distinct electrohydrodynamic flow fields generated during the UV ON/OFF states. On this basis, the proposed method allows the design of electrically driven microrobots with remote actuation and control. The ability to perform robust optical steering enables both open and closed-loop control schemes for particle navigation. The results conclusively demonstrate that closed-loop control extends the device's capability for precise maneuvering through a series of target points, ensuring robust and repeatable trajectory programming. Thus, our generic approach for optically gated steering of complex particles with photosensitive patches paves the way for designing future navigating controlled micromotors with broad versatility.

**Materials and methods**

*Microswimmers and photoconductivity test structures*:

The microswimmers' structure consists of a functional photoconductor and metal thin film selectively deposited over an epoxy body, fabricated on a silicon substrate. The photoconductivity test structures were fabricated on the same wafer using the same processes, to allow true measurement of the microswimmer robot photoconductive coating performance. The manufacturing process for the microswimmer is depicted in Figure 5, and for the photoconductivity test structure in supplementary Figure S2.

Initially, a sacrificial swimmer release layer (Kayaku LOR5A) was spin-coated onto a 4" single-side polished silicon wafer. Epoxy structures forming the base layer were then created on the surface using Gersteltec Sàrl GM1040 negative epoxy photoresist (Mask #1), after which the surface underwent 1 minute of Ar plasma treatment. Subsequently, a 200nm layer of ZnO was sputtered at room temperature to serve as the photoconductive layer. The ZnO layer was patterned using photolithography and etching processes (AZ 1505 positive photoresist, 0.3% volumetric HCl etch, Mask #2). A liftoff pattern was then deposited (AZ nLOF 2020, Mask #3), followed by sputtering of a 10nm Ti adhesion layer and 30nm of Au. A lift-off process was conducted using an Acetone ultrasonic bath. An additional lift-off mask (AZ 1518 photoresist, Mask #4) was applied to ensure good electrical contact with the test structure contact pads before passivation. Following 1 minute of $O_2$ plasma treatment, a 10nm sputter $SiO_2$ layer was deposited for surface passivation. The surface was protected for dicing with AZ photoresist 1518 and then diced to separate the microswimmers (5x5mm dies) and test structures (20x40mm). Photoconductivity test structures were cleaned with acetone and connected to electrodes using copper adhesive tape for testing (see Figure 2c inset). The particle dies underwent acetone cleaning to remove surface debris and were then agitated in NMP for particle release from the surface within a 1.5ml Eppendorf tube. The NMP solution in the tube was subsequently replaced with low conductivity KCl solution (3µS/cm) using a centrifuge (3 x 5-minute runs at 5kRPM).

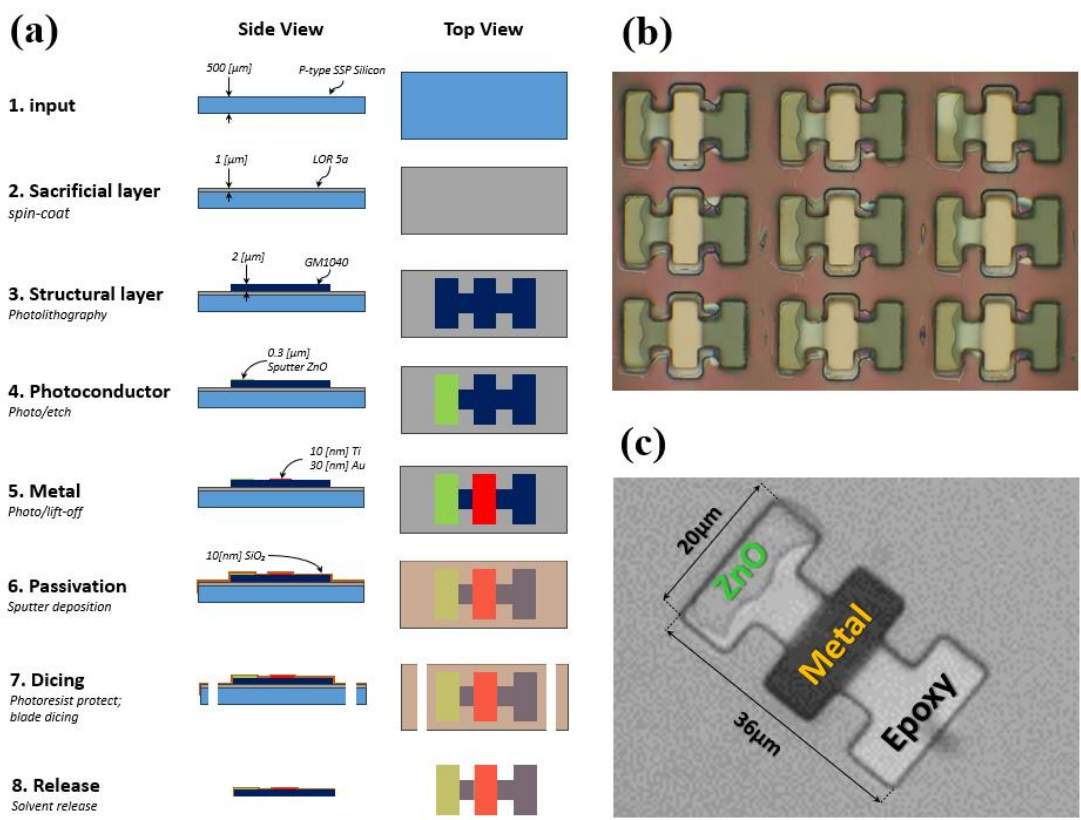

**Figure 5: Manufacturing of the Microswimmer robots.** (a) shows the manufacturing steps for the "HH" microswimmer robots. Microscope images show the microswimmer (b) on-wafer prior to release and (c) after release, suspended in fluid inside the test chamber.

*Test chamber:*

The test chamber consisted of two glass slides coated with Indium-Tin-Oxide (ITO) conductive coating at 4- Ω/sq (Delta Technologies, Cat. No. CB-40IN-0111). The bottom slide was sputter-coated with 10nm of SiO₂, while the top glass slide was drilled with 1mm diameter holes for inlet and outlet. The slides were separated by a 120μm adhesive sticker with an 8mm diameter circular cutout, effectively forming the chamber. The ITO-coated side of the slide was then connected to terminal leads using copper tape.

*Setup:*

The test chamber was placed atop a Nikon TI inverted microscope equipped with an Andor Neo camera, connected to an Agilent 33250A function generator for electrical signal generation, and a TPS 2024 Tektronix oscilloscope for signal monitoring (Figure s1). Illumination within the test chamber was provided by a top-side bright field light source, filtered using a green interference filter. UV activation light was introduced from below the sample using a Prizmatix Mic-LED-365L 365 nm LED light source. To facilitate activation of the photoconductor without excessive image saturation, the UV light passed through a 400 nm dichroic mirror. Both image and UV light were magnified using a Nikon Plan flour 20X/0.50 objective. The power density of the UV light was measured through an ITO-coated slide (Delta Technologies, Cat. No. CB-40IN-0111) to simulate test conditions, employing a 150μm diameter circular aperture. Signal measurement was conducted using an Ophir Photonics PD300 UV sensor connected to an Ophir Photonics VEGA power meter. Conductivity of the test structures was measured using a Keithley 2636A SMU.

*Microswimmer tracking and velocity measurements:*

We tracked the position, velocity and orientation of the microswimmer, both in real time during closed loop control, and offline to analyze the acquired data. We used Python to implement a tracking pipeline based on image processing methods, integrating functions from the open-source library OpenCV (See supplementary Figure S3 for details). Since the visual appearance of the swimmer changes drastically between the UV ON and UV OFF states, common solutions like the generic CSRT tracker fail to keep track of the swimmer. Furthermore, extracting orientation with such trackers is not straightforward. We therefore tailored a tracker for the microswimmer by performing separate processing sequences for each of the two modes. When the UV light is turned off, the swimmer appears dark on a slightly brighter background, and its edges are quite clear. For this mode we use Canny edge detection to isolate the edges, morphology operations to fill the gaps and clean noise, contour finding to find the largest surrounding contour, and minimum area rectangle fitting to represent the swimmer. The aspect ratio of the rectangle is also computed to determine if the swimmer is standing upright (high aspect ratio) or lying flat on the chamber floor (low aspect ratio). This adjustment is necessary when the swimmer either lifts from the chamber floor or falls from an upright position, as these actions cause a sudden and significant change in the tracked center position. Such a change is typically classified as a tracking error, causing the current frame to be ignored, but it is considered valid in these specific instances. When the UV light is turned on, the swimmer appears brighter on a bright background and its two exterior surfaces glow due to reflected light. For this mode we use binary thresholding to get an image in which the exteriors are clearly isolated from the background, followed by similar morphology operations and contour finding as the UV OFF case. The isolated contours are two blobs – one for each particle side. The blob centers of mass are connected by a line to find the middle point and extract orientation from the slope of the line. For both modes when the particle was moving, the orientation estimate was based on the angle of the velocity vector rather than the pure swimmer orientation for three reasons; First, the velocity direction is the control parameter for our closed loop motion control. Second, there is some misalignment between the swimmers' orientation and actual direction of motion. Third, the particle appears symmetric from top view most of the time, so there are always two solutions for orientation. To find the correct solution some memory-based information is needed, which can be unstable. To determine if the UV light was OFF or ON, we calculated the change in the mean value of the image's overall intensity compared to the previous frame. If this change was positive and exceeded a threshold value, it indicated that the UV light had switched from OFF to ON, and vice versa. We had to skip transition frames, as they typically did not fit any of the processing branches and could cause miscalculations. When a swimmer is first marked for tracking by manually placing a bounding box around it, the tracker defines a rectangular region of interest (ROI) around that bounding box. This ROI is cropped from the complete image to enable faster local processing and to avoid tracking other swimmers. As the tracked position is updated, the ROI is also updated and follows the tracked swimmer.

*Control system and algorithm:*

The control system was based on a custom-made Python program that controls the camera to get the image feed and communicates with the hardware by sending digital output commands to the electric function generator and optical illumination system. For the function generator, the controlled variables were voltage, frequency, and switching output ON/OFF. For the UV LED light source, the commands were setting power percentage and switching output ON/OFF. During live imaging of the sample the target microswimmer was selected and tracked, after a path planning module was executed to mark ordered waypoints to create the desired trajectory. The control pipeline operates on every new frame, so that the control rate can match the camera frame rate. However, to prevent very quick changes in hardware state, especially in UV on / off conditions, the control rate was limited to 5 Hz. The pipeline involves updating the tracking state, extracting position and orientation, comparing the current position of the swimmer with the current target waypoint, and applying the control algorithm to make a decision that will minimize the error and aim to reach that waypoint. When the swimmer reaches waypoint *n*, the focus switches to waypoint *n+1*, until it reaches the last one and automatically stops motion by setting the appropriate output. Since every 'HH' microswimmer can either move forward or rotate in only one direction – clockwise (CW) or counterclockwise (CCW), the controlled variables were reduced to implement a simple 'bang-bang' type controller. The controller switches the UV light on to turn the swimmer while moving, or off to move forward without turning. The electric field always stayed on with a constant predetermined voltage and frequency to maintain favorable ICEP propulsion conditions, and the UV power was set to 100% when turned on. Prior to a controlled navigation act, the microswimmer was put in an "idle" state by aligning the particle to an upright position. This is done by activating a low frequency electric field, and immediately after switching to high frequency (5 MHz) to keep it standing and immobile until navigation starts. The controller would automatically switch to 2 kHz to start moving, and back to 5 MHz upon finishing to return to the idle state and remain stationary. The control algorithm uses several constant preset parameters that can be tuned manually before a control session, and dynamic variables computed on every update. The first preset is the distance threshold $R$, which defines a radius around the target waypoint in which the swimmer is considered to have reached that target. Another essential preset is the activated turning direction of the swimmer (clockwise or counterclockwise), which is manually set by the operator for a specific navigation act. Controlled navigation begins with the microswimmer in an idle upright state to prevent it from switching turning directions between clockwise and counterclockwise, as it can align from flat position in 2 different chiral states, which are visually indistinguishable when upright. During navigation, the algorithm has three possible decisions to make due to the microswimmers single-sided turning capability. (1) move forward – if the swimmer is sufficiently aligned with the target waypoint, (2) turn - to correct alignment deviations in the activated turning direction, and (3) correction loop – execute a long rotation to produce the same effect as turning in the opposite direction. The error variable is defined as the deviation angle $\Delta\theta$ between the distance vector $\vec{d}$ to the target waypoint and the microswimmer orientation, which is determined from its velocity vector $\vec{v}$. The tolerance angle $T$ determines the maximum allowed $\Delta\theta$ in the activated turning direction before initiating a turn, and the correction gap $\phi$ is the maximum allowed $\Delta\theta$ in the opposite direction, beyond which a correction loop will be initiated. While the correction gap $\phi$ is a constant preset parameter, the tolerance $T$ is computed dynamically and is linearly proportional to the distance from the target $|\vec{d}|$ normalized by the maximum possible distance $d_{max}$, which is constant and determined from the swimmer size and image size in pixels. This computation is based on two additional preset parameters which are tunable – the tolerance near the target $T_N$ and the tolerance far from the target $T_F$, where $T_F > T_N > 0$, according to the following equation:

$$T(|\vec{d}|) = T_N + (T_F - T_N) \cdot \frac{|\vec{d}|}{d_{max}} \qquad (1)$$

The decision process (see scheme in Figure 6) for a swimmer is: (1) Update tracking and compute parameters. (2) check if arrived at the current waypoint ($|\vec{d}| < R$). If so, switch to the next waypoint (or finish if this is the last waypoint). Otherwise, continue. (3) Evaluate $\Delta\theta$; If $\Delta\theta > T$– perform an activated turn (turn UV on). If $\Delta\theta < -\phi$ – perform a correction loop (turn UV on). Else, keep moving forward (turn UV off). The sign of $\Delta\theta$ is positive if the deviation is in the same direction as the activated turn direction (up to 180°), and negative in the opposite case. One can notice that crossing 180° from the vector $\vec{d}$ will simply flip the sign of $\Delta\theta$ and switch the decision between correction loop and activated turn, which are realized in the same way, so there is a smooth transition in both directions.

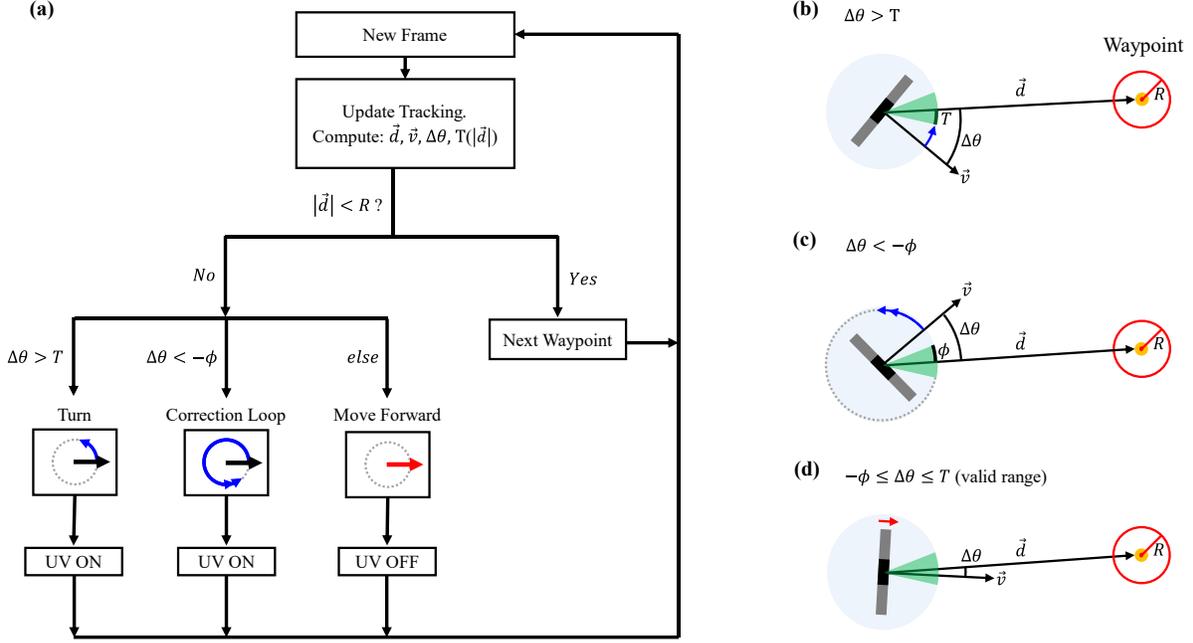

**Figure 6: Microswimmer closed-loop control scheme.** Shown for a counter-clockwise case - the same is valid for the clockwise case with flipped directions. (a) Flow chart of the algorithm used to decide whether to turn UV on / off. (b) The case where $\Delta\theta > T$ requires correcting the deviation with an active turn. (c) The case where $\Delta\theta < -\phi$ requires a correction loop. (d) When $\Delta\theta$ is in the valid range, the swimmer will continue forward.

*Numerical simulations:*

A three-dimensional numerical simulation was conducted using COMSOL™ 5.6 to qualitatively understand the electric potential and velocity fields surrounding an HH-shaped engineered particle under the influence of the ICEO mode. The geometries of the 'HH' structures which uniformly matched the fabricated particle, were positioned 1 µm above the substrates of cubic microchamber domains (100 µm height, 500 µm length and depth). For the electrostatic analysis, the Laplace equation was solved for the electric potential $\phi$, in conjunction with the boundary condition for the formation of the induced electric double layer (EDL) on the gold and ZnO (only under UV ON state) surfaces, assuming a thin EDL over the length of the engineered particle. $\left(\frac{\lambda}{l} \ll 1\right)$(23,24),

$$\sigma \frac{\partial \tilde{\phi}}{\partial n} = i\omega C_{DL}(\tilde{\phi} - V_{Au}), \quad \sigma \frac{\partial \tilde{\phi}}{\partial n} = i\omega C_{DL}(\tilde{\phi} - V_{ZnO}) \qquad (2)$$

wherein $\tilde{\phi}$ is the phasor of the (complex) electric potential, $V_{Au}$ and $V_{ZnO}$ are the floating potentials of the gold and ZnO surface of the engineered particle respectively, $n$ is the normal coordinate to the particle surface, and $C_{DL}$ represents the capacitance per unit area of the EDL which may be also simply estimated from the Debye-Huckel theory as $C_{DL} \sim \varepsilon/\lambda$ (neglecting the capacitance of the Stern layer). The external electric potentials at the lower substrate (z = 0) and on the upper wall of the channel were set to 7.5V and -7.5V, respectively, while insulating boundary conditions were applied to the other surfaces of the dielectric part (epoxy) and the sides of the microchamber. For qualitative purposes, we neglected the induced EDL screening of the powered electrodes, which is only influential at frequencies much smaller than that related to the RC time of the EDL induced on the floating gold and ZnO surfaces. Additionally at the UV Off state, the electrostatic boundary condition of the ZnO surface was set to an insulating surface.

The hydrodynamic equations, under the assumption of weak-field, small Peclet number and thin EDL, can be decoupled from the electrostatic equation and used to obtain the velocity field via the unforced Stokes equation. Electric forcing was incorporated by prescribing an effective Helmholtz-Smoluchowski (HS) slip velocity both on the Au and ZnO (only under UV ON state) surfaces. The time-averaged HS slip velocity boundary condition on these surfaces is (25,26)

$$u_S = \frac{1}{2}\frac{\varepsilon}{\eta} Re\big(\tilde{\zeta}\tilde{E}_\parallel^*\big) \qquad (3)$$

where $\tilde{\zeta}$ is the (complex) induced surface zeta potential, which is equal to either $(V_{Au} - \phi)$ or $(V_{ZnO} - \phi)$, and $\tilde{E}_\parallel^*$ is the complex conjugate of the tangential surface component of the electric field $\tilde{\mathbf{E}} = -\nabla\tilde{\phi}$, and $\eta$ is the dynamical viscosity of solution. On all other walls, no-slip and no-penetration conditions were applied. Also under no UV activation, boundary condition on the ZnO surface was set to no-slip.

**Acknowledgements**

G.Y. acknowledges support from the Israel Science Foundation (ISF) (1934/20). M.Z., T.M., O.D.V. and G.Y. acknowledge support from the Binational Science Foundation (BSF) grant 2018168. We would like to thank the nanocenter in Tel-Aviv University for assisting in fabrication of the micromotors.

**Acknowledgements**

We thank the Tel-Aviv University Nanoscience and Nanotechnology Center for their assistance in fabricating the particles and chip. **Funding:** G.Y. acknowledges support from the Israel Science Foundation (ISF) (1934/20). M.Z., T.M., O.D.V. and G.Y. acknowledge support from the Binational Science Foundation (BSF) grant 2018168. We would like to thank the nanocenter in Tel-Aviv University for assisting in fabrication of the micromotors. Author contributions: G.Y., O.V., and T.M. conceived the concept of the optically-gated micromotor. M.Z. and G.Y. designed the mechanism. G.Y. supervised the project. M.Z. led the experiments, data analysis, and manuscript writing. I.R. and M.Z. developed the closed-loop system and conducted the experiments. S.P. performed the numerical simulations and assisted M.Z. with the micromotor fabrication. All authors contributed to the manuscript writing. **Competing interests:** The authors declare no competing financial interests. **Data and materials availability:** All data needed to evaluate the conclusions in the paper are present in the paper or the Supplementary Materials.

**Supplementary Materials**

# Programmable Motion of Optically Gated Electrically Powered Engineered Microswimmer Robots

Matan Zehavi[1], Ido Rachbuch[2], Sinwook Park[2], Touvia Miloh[2], Orlin D. Velev[3] and Gilad Yossifon[2,4*]

[1]*Faculty of Mechanical Engineering, Technion–Israel Institute of Technology, Technion City 32000, Israel*
[2]*School of Mechanical Engineering, Tel Aviv University Ramat Aviv 69978, Israel*
[3]*Department of Chemical and Biomolecular Engineering, NC State University Raleigh, NC 27695, USA*
[4]*Department of Biomedical Engineering, Tel Aviv University Ramat Aviv 69978, Israel*

* Corresponding author: gyossifon@tauex.ta.ac.il

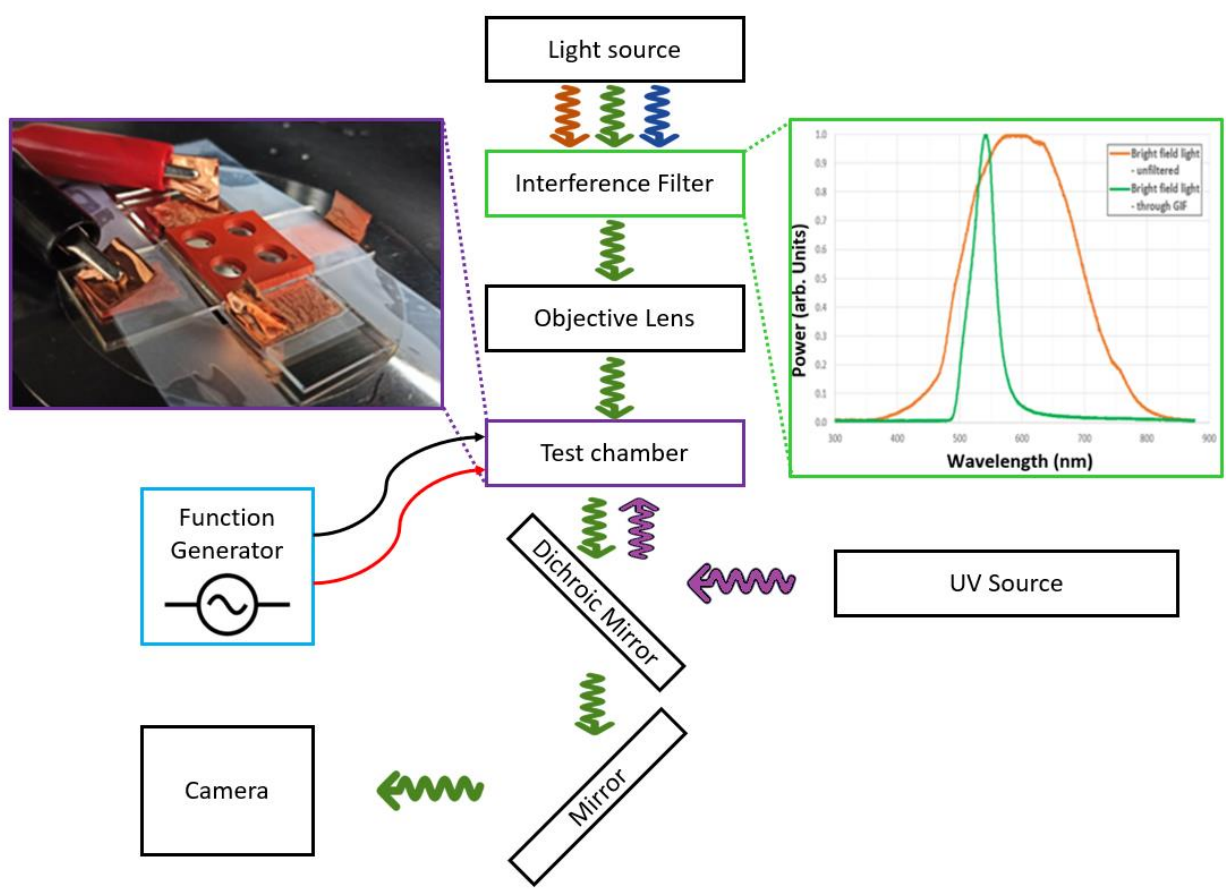

**Figure S1: Testing setup for microswimmer experiments.** A block diagram of the critical components for the microswimmer experiments. The right inset shows the wavelength profile of the source and filtered imaging flood light.

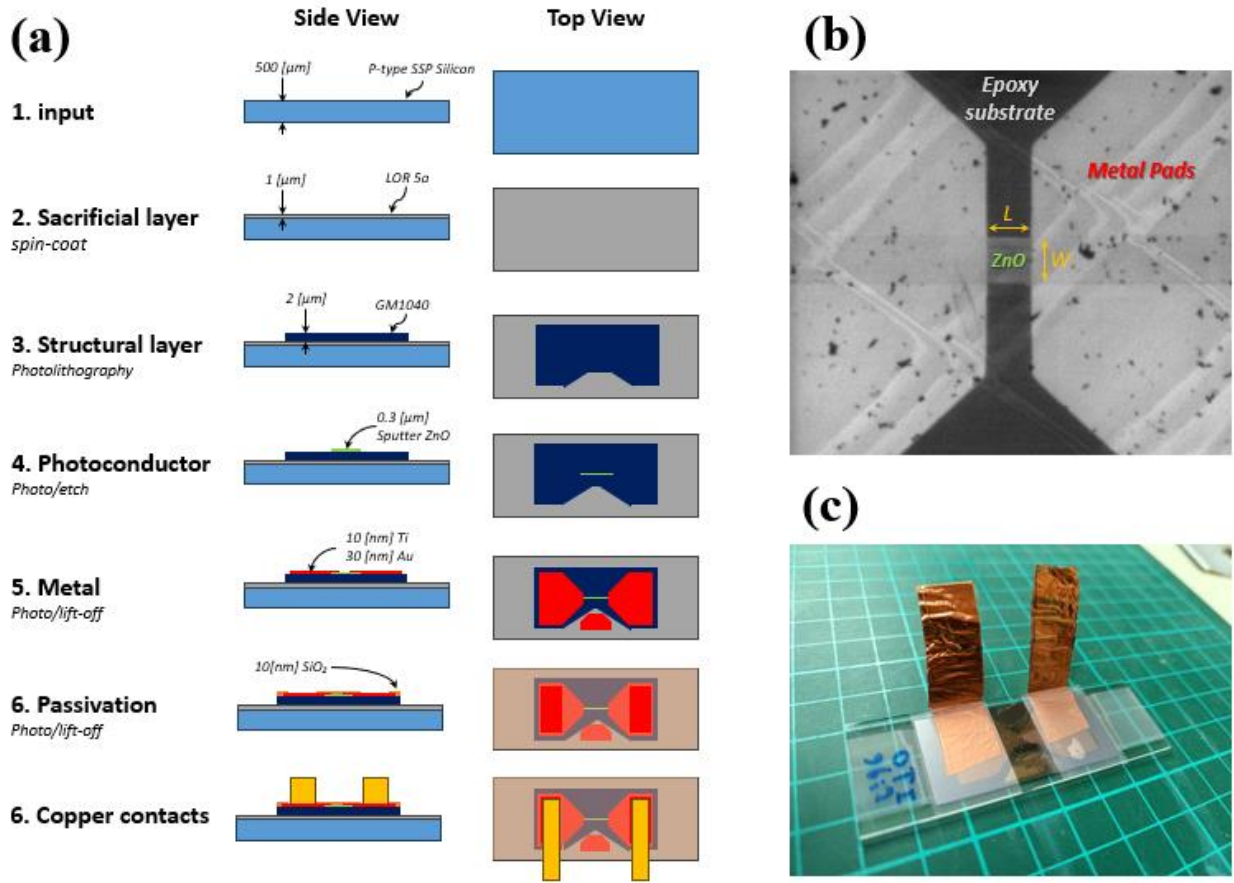

**Figure S2: Photoconductivity test structure**. (a) manufacturing scheme. (b) microscope image of the ZnO active area. Two different structures were designed and manufactured with varying LxW designs: 100 µm x100µm and 200 µm x200µm. Thickness was determined by ZnO deposition. (c) Two assembled test structures.

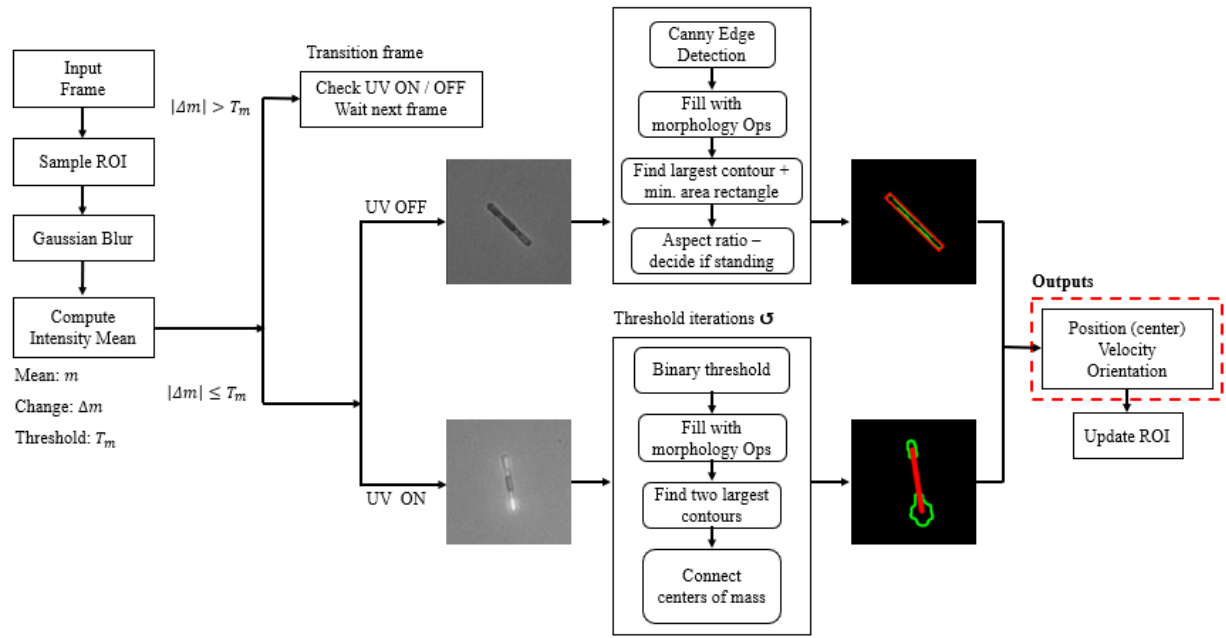

**Figure S3: Microswimmer tracking pipeline.** First, the relevant ROI around the microswimmer is cropped from the input frame. Gaussian blur is then applied to reduce noise, and the intensity mean of the image is calculated to compare to the previous frame and compute the change: $\Delta m$. If $\Delta m$ is positive and larger than the threshold value $T_m$, it means that there has been a transition from the UV OFF to the UV ON state, and vice versa. In this case, the frame is considered a "transition frame" and skipped, while subsequent frames are processed according to the new state. Each frame follows one of two sequences tailored to the microswimmer's appearance in the given state. In the UV OFF case, the borders of the swimmer are isolated with Canny edge detection. In the UV ON state, the swimmer's two glowing edges are isolated using binary thresholding, with several iterations at lower thresholds until they are identified. The extracted outputs include position, velocity and orientation. The new position is used to update the ROI for the next frame.

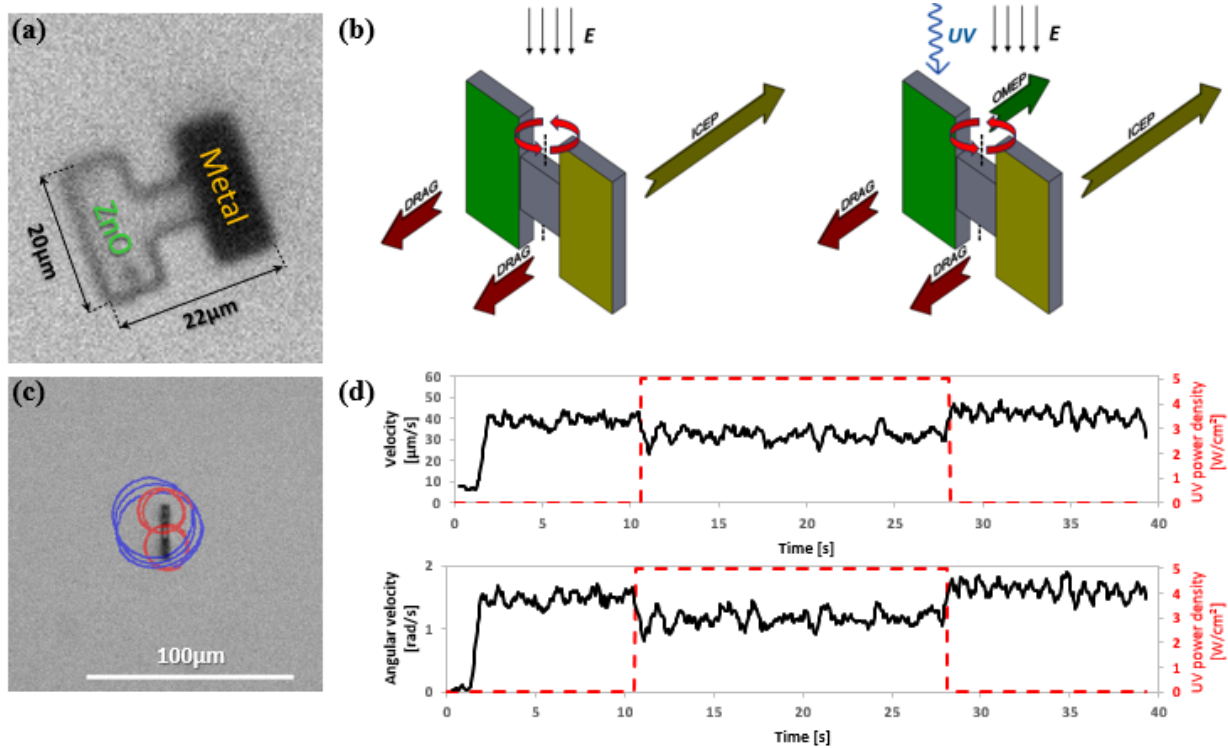

**Figure S4: 'H'- type particle response.** (a) Microscope image of the 'H'-shaped microswimmer. (b) Operational principle of steering under an externally applied electric field. (c) Trajectory of the microswimmer, and corresponding (d) linear and angular velocities recorded in an open-loop control scenario with UV switching on and off, under an AC electric field of 2 kHz and 1666 V/cm (see supplementary video SV5).

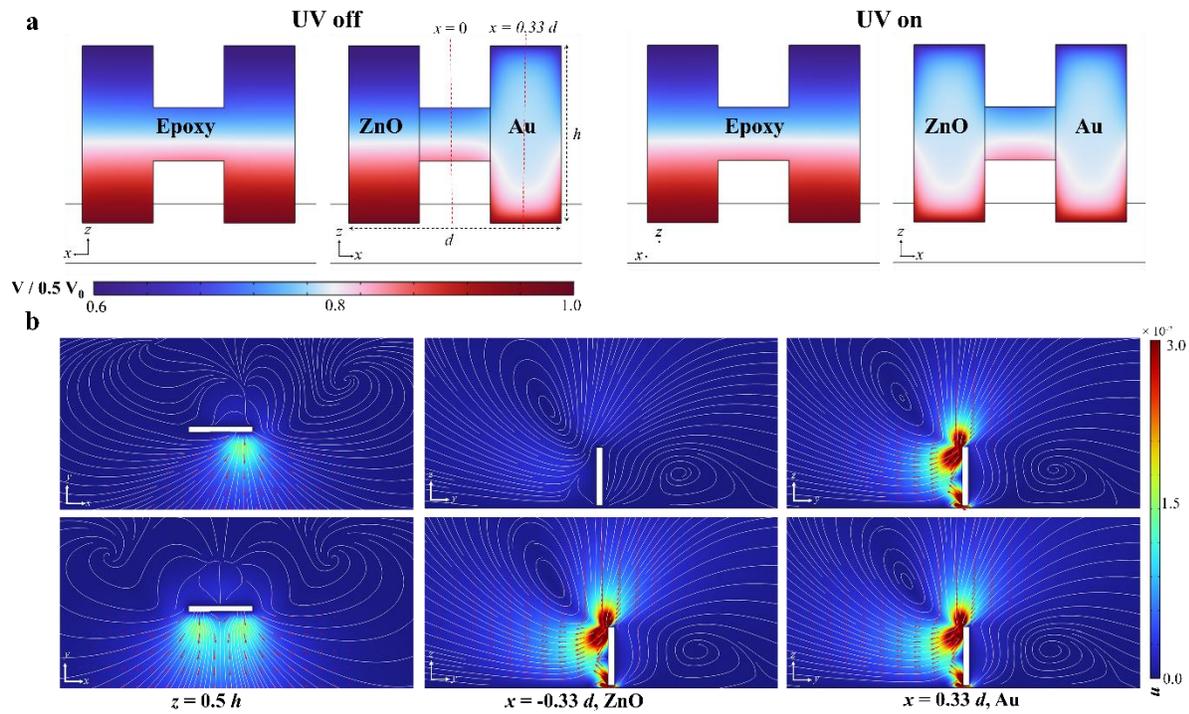

**Figure S5: Numerical simulation of 'H'- type particle.** (a) electric field intensity simulations results of front (Metal and ZnO) and back (Epoxy) views of the 'H' type swimmer, showcasing normalized electric potential intensities in UV ON and OFF states. (b) Velocity field and streamline plots at the particle surface during both operational states, taken from various plane views. Magnitude is indicated by color.